\documentclass[lettersize,journal]{IEEEtran}
\usepackage{amsmath,amsfonts}
\usepackage{algorithmic}
\usepackage{algorithm}
\usepackage{array}
\usepackage{booktabs}
\usepackage{threeparttable}
\usepackage[caption=false,font=normalsize,labelfont=sf,textfont=sf]{subfig}
\usepackage{textcomp}
\usepackage{stfloats}
\usepackage{url}
\usepackage{verbatim}
\usepackage{graphicx}
\usepackage{cite}
\newcolumntype{L}[1]{>{\raggedright\arraybackslash}p{#1}}
\newcolumntype{C}[1]{>{\centering\arraybackslash}p{#1}}

\begin{document}

\title{Demonstration and Design of Uni-Directional 
and Ultra-Low Threshold Hybrid Quantum Dot III-V/Si 
Micro-Ring Lasers}

\author{Xucheng Yang, Yingtao Hu, Antoine Descos, Yuan Yuan, Bassem Tossoun, Geza Kurczveil, Yatiraj Ramanujam, Jonathan Wierer, Raymond G. Beausoleil, Di Liang, Stanley Cheung
\thanks{This work was supported in part by the Advanced Research Projects Agency–Energy under Grant DE-AR0001039.}
\thanks{Xucheng Yang, Jonathan Wierer and Stanley Cheung are with the North Carolina State University, Dept. of Electrical and Computer Eng., Raleigh, NC. 27606, USA.

Antoine Descos, Yingtao Hu, Bassem Tossoun, Geza Kurczveil and Raymond G. Beausoleil are with the Hewlett Packard Labs, Large-Scale Integrated Photonics Laboratory, Milpitas, CA. 95035, USA.

Yatiraj Ramanujam and Di Liang are with the University of Michigan, Ann Arbor, MI. 48109, USA.

Yuan Yuan is with the Northeastern University, Department of Electrical Engineering, Oakland, CA. 94613, USA.}}

\markboth{Journal of \LaTeX\ Class Files,~Vol.~14, No.~8, August~2021}%
{Shell \MakeLowercase{\textit{et al.}}: A Sample Article Using IEEEtran.cls for IEEE Journals}


\maketitle

\begin{abstract}
Micro-ring lasers (MRLs) are attractive light sources for energy-efficient optical interconnects, but their intrinsic directional bistability leads to unpredictable clockwise/counter-clockwise emission. We demonstrate stable unidirectional emission in hybrid quantum-dot (QD) III--V/Si MRLs using passive reflective feedback integrated on the bus waveguide, leaving the ring cavity unperturbed. Three reflector architectures---Y-splitter loop mirrors, adiabatic Y-splitter loop mirrors, and distributed Bragg reflectors (DBRs)---are benchmarked against a reflector-free bidirectional baseline through combined experiment and coupled-mode-theory rate-equation modeling. All designs preserve ultra-low thresholds of 0.79--1.12~mA (112--158~A/cm\textsuperscript{2}, roughly an order of magnitude below prior quantum-well unidirectional ring lasers) while enhancing single-facet output power and wall-plug efficiency, with directional isolation up to 27.65~dB for the DBR. The reflectors impose no penalty on the 4--5~GHz modulation bandwidth or its thermal robustness, establishing passive external feedback as a practical route to unidirectional QD MRLs for DWDM-scale optical interconnects.
\end{abstract}


\section{Introduction}
\IEEEPARstart{M}{icro-ring} lasers (MRLs) have emerged as an important on-chip light-source architecture for optical interconnects because they combine compact footprint, high integration density, wavelength-division multiplexing scalability\cite{ref1,ref2,ref3}, and direct modulation capability within a resonant cavity platform\cite{ref4,ref5}. The resonant enhancement of circulating optical power in the ring cavity enables low-threshold and low-energy operation\cite{ref6,ref7}, making MRLs highly attractive for short-reach data links, photonic co-packaging, and other energy-efficient integrated photonic systems\cite{ref8,ref9,ref10,ref11,ref12,ref13}. However, MRLs intrinsically exhibit directional bistability, meaning that lasing can occur in either the clockwise (CW) or counter-clockwise (CCW) direction\cite{ref14,ref15}. Although this property has been utilized for optical switching and photonic logic functionalities\cite{ref16,ref17}, it is undesirable for optical interconnects because it causes unpredictable coupling into downstream components. As a result, stable unidirectional lasing is highly desirable for practical MRL-based interconnect systems. Several approaches have been explored to enforce directional or modal selectivity in MRLs. One representative strategy is to introduce intracavity symmetry-breaking structures, such as an S-bend waveguide that enables asymmetric coupling between the counter-propagating modes\cite{ref18}, or azimuthal gratings integrated along the ring perimeter to select a specific mode\cite{ref19,ref20}, a representative layout of which is shown in Fig.~\ref{fig4}(d). While these methods can effectively favor one lasing direction or suppress competing modes, they directly perturb the cavity itself and are therefore susceptible to fabrication uncertainty and additional scattering or radiation loss. As a result, the cavity quality factor can be degraded, which in turn leads to increased threshold current and reduced slope efficiency. Another route is to enhance the net gain in one direction through external optical injection through external laser or LED\cite{ref21,ref22}. However, such approaches generally increase system complexity and power consumption.

Despite rapid progress in heterogeneously integrated III--V/Si QD microring lasers\cite{ref4,ref7,ref12,ref23}, passive external feedback for unidirectional operation has been demonstrated primarily on conventional quantum-well hybrid-silicon platforms\cite{ref24,ref25}, and the available reflector options have not been compared on a common QD platform. As summarized in Table~\ref{tab:benchmark}, prior unidirectional ring and disk lasers---realized on quantum-well gain media through either intracavity symmetry-breaking or external-feedback schemes---exhibit threshold current densities on the order of $1$--$2~\mathrm{kA/cm^2}$. How the reflector architecture should be co-optimized with a quantum-dot gain medium to enforce single-ended emission while preserving its intrinsic ultra-low threshold and high-speed capability thus remains largely open. Bridging reflector design, coupling conditions, and QD device physics, this work demonstrates stable single-ended emission at ultra-low threshold currents of approximately 1~mA (0.79--1.12~mA across the three designs), corresponding to threshold current densities of only $\sim$112--158~A/cm$^2$---roughly an order of magnitude below those prior demonstrations (Table~\ref{tab:benchmark})---with single-facet output power and wall-plug efficiency exceeding those of the bidirectional baseline, a directional isolation of up to 27.65~dB for the DBR, and no measurable penalty on the modulation bandwidth ($\sim$4--5~GHz). Benchmarking three passive feedback architectures---Y-splitter loop mirrors, adiabatic Y-splitter loop mirrors, and DBRs---against a reflector-free baseline through combined experiment and modeling, we identify the key trade-offs among threshold, output power, efficiency, directional isolation, device footprint, and high-speed response, and provide design guidelines for scalable, energy-efficient unidirectional on-chip laser integration.

\begin{table*}[!t]
\centering
\caption{State-of-the-art uni-directional III-V/Si microring/microdisk lasers}
\label{tab:benchmark}
\renewcommand{\arraystretch}{1.25}
\setlength{\tabcolsep}{4pt}
\begin{threeparttable}
\begin{tabular}{L{0.2\textwidth} L{0.16\textwidth} L{0.21\textwidth} C{0.10\textwidth} C{0.18\textwidth}}
\toprule
Reference & Active region & Unidirectional mechanism & $J_{\mathrm{th}}$ (A\,cm$^{-2}$) & Directional isolation (dB) \\
\midrule
Liang \emph{et al.}~\cite{ref24} & InAlGaAs MQW & External teardrop reflector & $\sim$2150 & single-ended\tnote{a} \\

Mechet \emph{et al.}~\cite{ref25} & InP MQW& External DBR & $\sim$1360 & 39.9--46.1 \\

This work (Y-splitter) & InAs/GaAs QD & External Y-splitter loop mirror & 140 & single-ended\tnote{a} \\

This work (adiabatic) &  InAs/GaAs QD & External adiabatic loop mirror & 158 & single-ended\tnote{a} \\

This work (DBR) & InAs/GaAs QD & External DBR & 112 & 27.65 \\
\bottomrule
\end{tabular}
\begin{tablenotes}[flushleft]
\footnotesize
\item[a] Loop-mirror devices have a single output facet, so a finite directional ratio is not defined.
\item For reference, the reflector-free QD baseline of this work has $J_{\mathrm{th}}=113$~A\,cm$^{-2}$ and $P_{\mathrm{CW}}/P_{\mathrm{CCW}}=-1.92$~dB.
\end{tablenotes}
\end{threeparttable}
\end{table*}

\section{Device design and operating principle}
\subsection{Baseline hybrid QD III-V/Si microring laser design}

\begin{figure}[!t]
\centering
\includegraphics[width=3.2in]{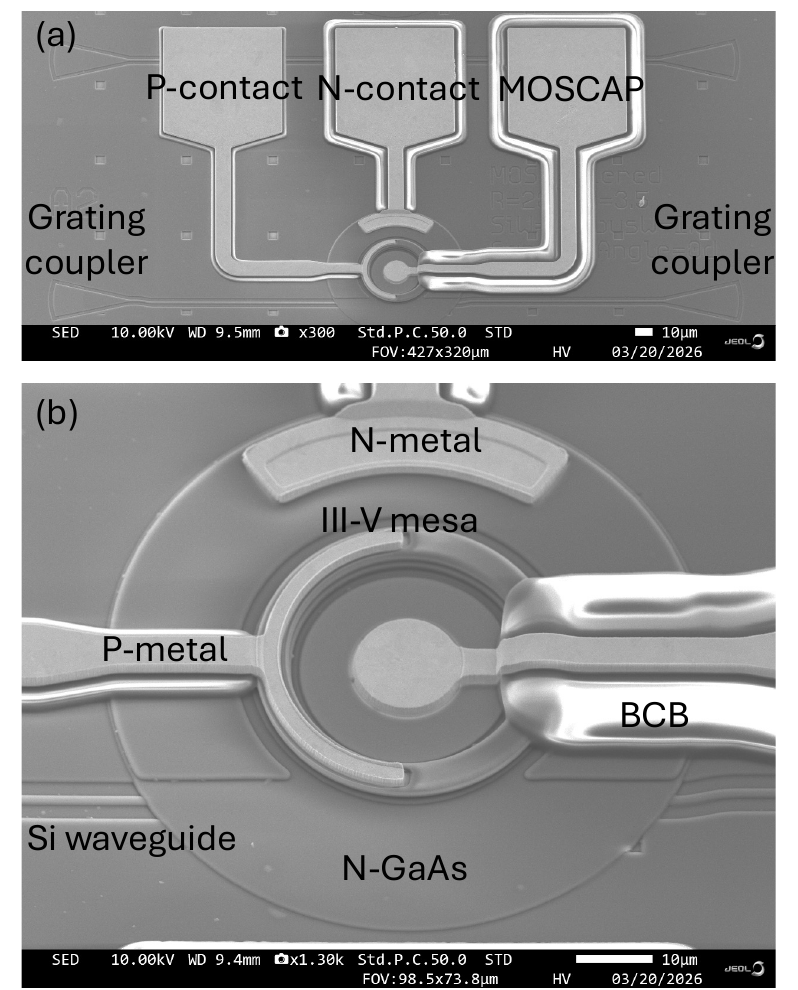}
\caption{(a) SEM image of fabricated hybrid III-V/Si 5QD MRL without external reflectors. (b) Close-up of active region.}
\label{fig1}
\end{figure}

\begin{figure}[!t]
\centering
\includegraphics[width=3in]{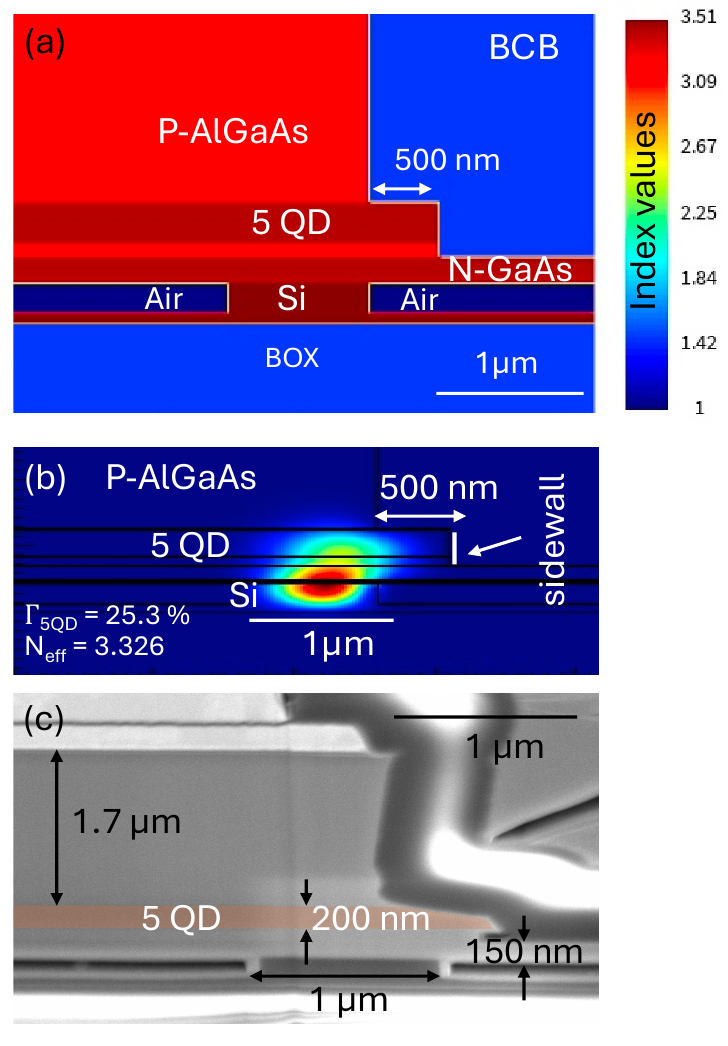}
\caption{(a) Cross-sectional design and refractive index values for the hybrid III-V/Si QD structure. (b) Mode simulation for silicon waveguide widths of $1.0~\mu\text{m}$. (c) Cross-sectional SEM and relevant dimensions.}
\label{fig2}
\end{figure}

As shown in Fig.\ref{fig1}(a)--(b), the baseline device is a reflector-free hybrid III--V/Si 5QD microring laser consisting of a $25~\mu\text{m}$-radius electrically pumped ring cavity side-coupled to a silicon bus waveguide, with optical outputs collected from the left and right grating-coupler ports. The reference cavity employs a $5~\mu\text{m}$-wide wafer-bonded III--V mesa with a p-i-n epitaxial stack and a five-layer InAs/GaAs quantum-dot active region formed on top of a $1.0~\mu\text{m}$-wide silicon ring waveguide. The silicon waveguide is defined by a height of $300~\text{nm}$ and an etch depth of $217~\text{nm}$ on the BOX layer, while the coupled bus waveguide is also $1.0~\mu\text{m}$ wide. The ring and bus are connected through a symmetric point-coupling region ($\theta_{\mathrm{coupler}} = 0^\circ$) with a $0.2~\mu\text{m}$ coupling gap, providing the nominal baseline coupling condition before any directional reflector is introduced. For this geometry, the fundamental hybrid mode, solved by Finite-Difference-Eigenmode (FDE), exhibits a QD optical confinement factor of $\Gamma_{\mathrm{QD}} = 25.3\%$ and an effective index of $n_{\mathrm{eff}} = 3.326$, as illustrated by the cross-sectional design and modal profile in Fig.\ref{fig2}(a)--(b). The mode is jointly confined in the QD gain region and the silicon waveguide, providing sufficient overlap for optical gain while remaining displaced from the etched mesa sidewall to reduce sidewall-induced scattering and surface recombination. In the absence of any external reflector, this device operates as the intrinsic bidirectional MRL reference, serving as the baseline platform for evaluating how external feedback modifies the emission directionality without altering the core ring cavity itself.

\subsection{Analysis of unidirectional lasing enabled by reflective feedback}
To capture the directional symmetry breaking induced by the external reflector, we model the hybrid III--V/Si QD microring laser using an extended coupled-mode-theory (CMT) rate-equation framework\cite{ref26}. The cavity supports two counter-propagating traveling-wave modes, namely the clockwise (CW) and counter-clockwise (CCW) fields, with complex slowly varying amplitudes $E_{\mathrm{CW}}$ and $E_{\mathrm{CCW}}$, respectively, where $|E|^2$ denotes the intracavity photon density (m$^{-3}$). Both modes share a common carrier reservoir with carrier density $N$. The field evolution is written as
\begin{align}
\frac{dE_{\mathrm{CW}}}{dt}
&=
\frac{1}{2}(1+j\alpha_H)\left(G^{+}-\frac{1}{\tau_p}\right)E_{\mathrm{CW}}
+jK_1E_{\mathrm{CCW}}, \label{eq:cw}\\
\frac{dE_{\mathrm{CCW}}}{dt}
&=
\frac{1}{2}(1+j\alpha_H)\left(G^{-}-\frac{1}{\tau_p}\right)E_{\mathrm{CCW}}
+jK_2E_{\mathrm{CW}}, \label{eq:ccw}
\end{align}
where $\tau_p$ is the photon lifetime, $\alpha_H$ is the linewidth-enhancement factor, and $K_1$ and $K_2$ are the linear coupling coefficients between the two counter-propagating waves. The first term in each equation represents the net modal gain, while the last term describes the coherent backscattering that couples the CW and CCW fields.

For the baseline reflector-free cavity, the two coupling coefficients are equal, $K_1 = K_2 = K_b$, where $K_b$ is the intrinsic backscattering rate arising from residual sidewall roughness and the point-coupling region. Introducing the external reflector at the end of the bus waveguide breaks this symmetry and renders the coupling asymmetric,
\begin{equation}
K_1 = K_2 + |k|^2 r_1 \frac{v_g}{\pi D},
\label{eq:asym}
\end{equation}
where $K_2$ represents the intrinsic backscattering rate, $|k|^2$ is the power coupling coefficient between the ring and bus waveguide, $r_1$ is the amplitude reflectivity of the external reflector, $v_g$ is the group velocity, and $\pi D$ is the ring circumference. In this form, the reflector does not simply add gain or loss; instead, it breaks the CW/CCW symmetry by modifying the intermodal coupling asymmetrically, thereby biasing the cavity toward one circulation direction.

Above threshold, the final directional state is determined by nonlinear gain competition. The direction-dependent gains are written as
\begin{equation}
G^{\pm}
=
\frac{\Gamma g_0 v_g \ln(N/N_{\mathrm{tr}})}
{1+\varepsilon_s|E^{\pm}|^2+\varepsilon_c|E^{\mp}|^2},
\label{eq:gain}
\end{equation}
where $\Gamma$ is the optical confinement factor, $g_0$ is the gain coefficient of the QD active material\cite{ref27}, $N_{\mathrm{tr}}$ is the transparency carrier density, and $\varepsilon_s$ and $\varepsilon_c$ are the self- and cross-gain saturation coefficients with $\varepsilon_c = 2\varepsilon_s$\cite{ref30,ref31}, respectively. The carrier density evolves according to
\begin{equation}
\frac{dN}{dt}
=
\frac{\eta_i I}{qV}
-BN^2
-CN^3
-G^{+}|E_{\mathrm{CW}}|^2
-G^{-}|E_{\mathrm{CCW}}|^2,
\label{eq:carrier}
\end{equation}
where $\eta_i$ is the injection efficiency, $V$ is the QD gain volume, and $B$ and $C$ are the bimolecular and Auger recombination coefficients. A stochastic Gaussian noise term is added to each field equation at every time step to emulate the random spontaneous-emission process coupled into the CW and CCW modes. The coupled state vector $[E_{\mathrm{CW}},E_{\mathrm{CCW}},N]$ is then integrated in the time domain using a fixed-step fourth-order Runge--Kutta scheme. Table~\ref{tab:cmt_parameters} summarizes the parameters implemented in the numerical solution of the above set of equations.

\begin{figure}[!t]
\centering
\includegraphics[width=3in]{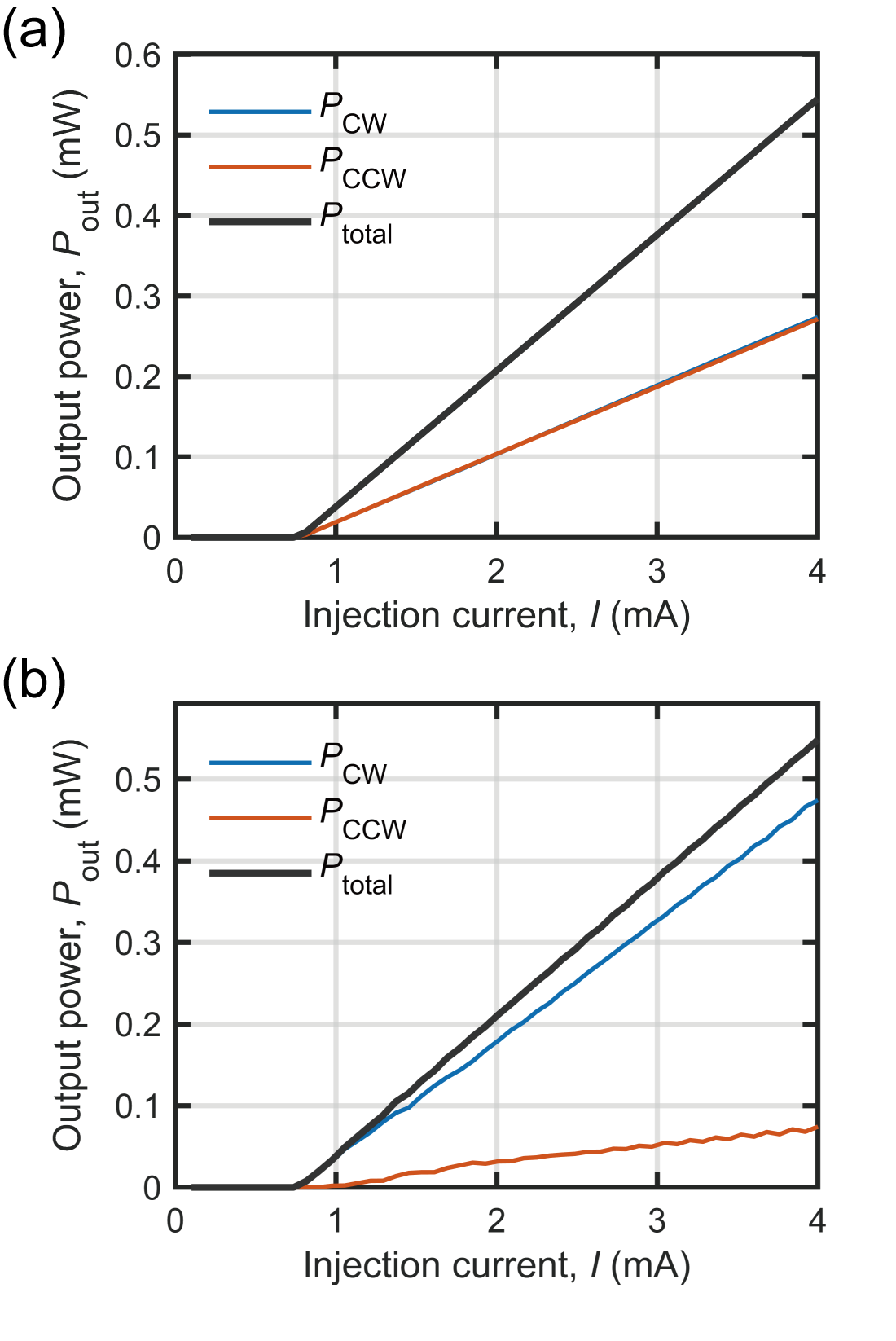}
\caption{Simulated steady-state light-current characteristics of (a) reflector-free and (b) reflector-assisted microring laser based on the extended CMT rate-equation model.}
\label{fig3}
\end{figure}

\begin{table}[t]
\centering
\caption{Fixed parameters used in the extended CMT rate-equation simulations.}
\label{tab:cmt_parameters}
\footnotesize
\setlength{\tabcolsep}{3pt}
\renewcommand{\arraystretch}{1.12}
\begin{tabular}{C{0.16\columnwidth} L{0.41\columnwidth} C{0.14\columnwidth} C{0.08\columnwidth} C{0.09\columnwidth}}
\toprule
Symbol & Description & Value & Unit & Source \\
\midrule
$v_g$ & Group velocity $(c/n_g)$ & $8.24 \times 10^{7}$ & m/s & \\
$g_0$ & Logarithmic gain coefficient & $3 \times 10^{5}$ & m$^{-1}$ & \cite{ref27} \\
$N_{\mathrm{tr}}$ & Transparency carrier density & $1.2 \times 10^{24}$ & m$^{-3}$ & \cite{ref28} \\
$\alpha_H$ & Linewidth enhancement factor & $1.0$ & -- & \\
$\varepsilon_s$ & Self-gain saturation & $1 \times 10^{-25}$ & m$^{3}$ & \\
$\varepsilon_c$ & Cross-gain saturation $(2\varepsilon_s)$ & $2 \times 10^{-25}$ & m$^{3}$ &  \\
$B$ & Bimolecular recombination & $1 \times 10^{-16}$ & m$^{3}$/s & \cite{ref28} \\
$C$ & Auger recombination & $3 \times 10^{-41}$ & m$^{6}$/s & \cite{ref29} \\
$\eta_i$ & Injection efficiency & $0.5$ & -- & \\
$\tau_p$ & Photon lifetime & $4.68$ & ps & \cite{ref27} \\
$r_1$ & External reflector reflectivity & $0.9$ & -- & \\
$|K_2|$ & Intrinsic backscattering rate & $5 \times 10^{8}$ & rad/s & \\
$|K_1 - K_2|$ & External reflector coupling contribution & $2.83 \times 10^{10}$ & rad/s & \\
\bottomrule
\end{tabular}
\end{table}

Fig.\ref{fig3} compares the simulated steady-state LI curves with and without the external reflector. In the reflector-free case, the CW and CCW output powers remain nearly identical above threshold. This is consistent with the nearly symmetric condition $K_1 \approx K_2$, for which the lasing power is distributed almost equally between the two directions. In contrast, when the external reflector is included, the total output power follows a similar overall current dependence, but the directional power balance becomes strongly asymmetric: one direction is preferentially enhanced, while the opposite direction is substantially suppressed. Therefore, the primary role of the external reflector is to redistribute the lasing power into a directionally preferred steady state through asymmetric modal coupling and nonlinear gain competition, rather than merely reflecting the output from one side of the bus waveguide to the other.

\subsection{Loop mirror design}

\begin{figure*}[!t]
\centering
\includegraphics[width=7.0in]{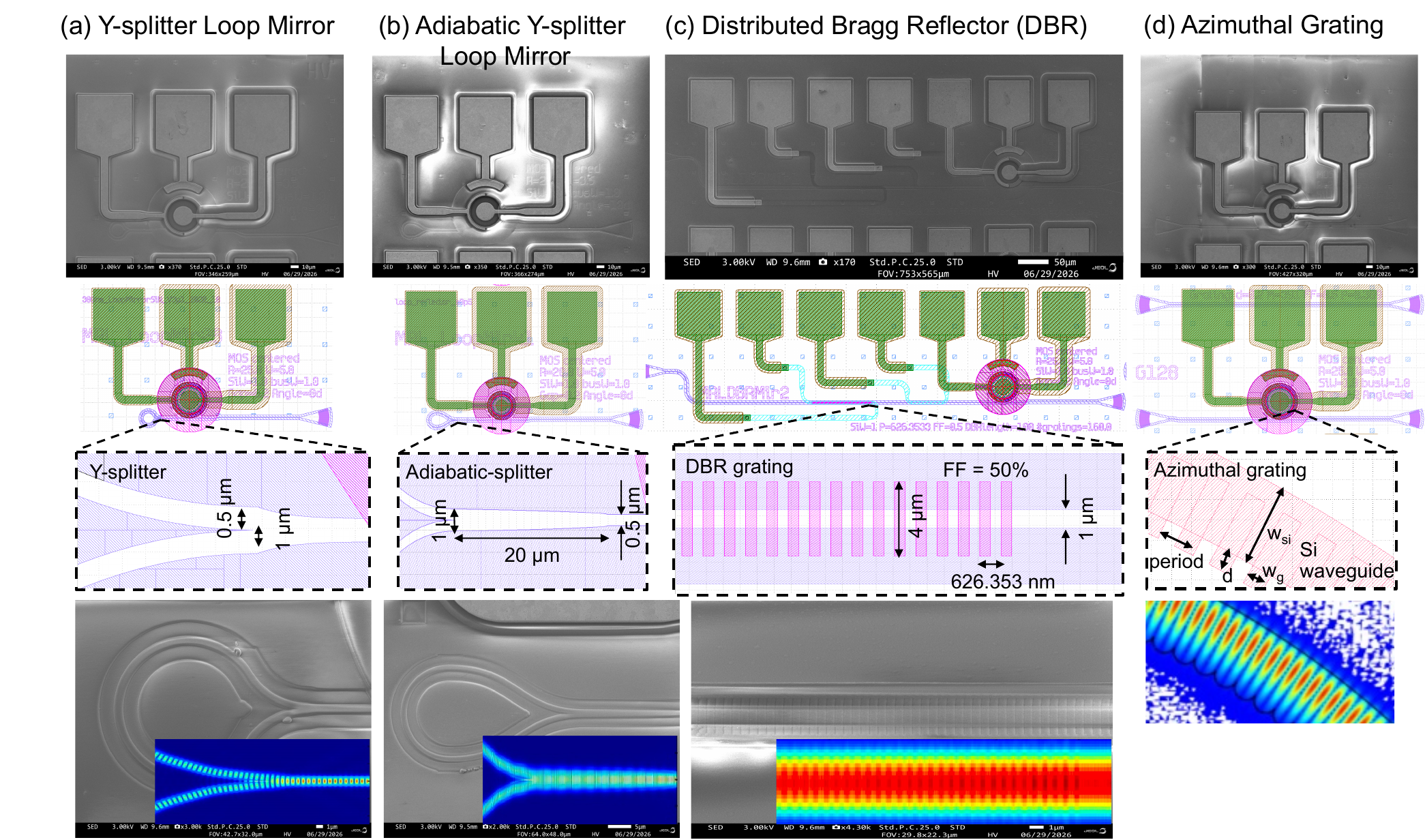}
\caption{SEM images (top row), mask layouts (second row), magnified schematics of the coupling/grating regions with key dimensions (third row), and FDTD-simulated field distributions (bottom row) for the various designs: (a) Y-splitter loop mirror, (b) adiabatic Y-splitter loop mirror, (c) DBR, and (d) intracavity azimuthal grating.}
\label{fig4}
\end{figure*}

To implement passive direction-selective feedback without directly modifying the microring cavity, loop mirrors were integrated at the end of the bus waveguide as shown in Fig.\ref{fig4}(a). The conventional Y-splitter loop mirror divides the incident field from the $1~\mu\mathrm{m}$-wide input waveguide into two $0.5~\mu\mathrm{m}$ arms at a relatively short junction; the two arms are then routed through a loop and recombined to reflect light back toward the bus. In this geometry, the modal transition at the junction is not fully adiabatic, which can lead to partial coupling of the incident field into higher-order or radiation modes and thus contribute to reduced reflectivity. In contrast, Fig.\ref{fig4}(b) shows an adiabatic splitter loop mirror, in which the $1~\mu\mathrm{m}$ bus waveguide gradually expands and branches into two $0.5~\mu\mathrm{m}$ arms over a $20~\mu\mathrm{m}$ transition region. The smoother modal evolution mitigates the mode mismatch present in the conventional Y-junction, tending to yield a more balanced 50:50 split, improved fabrication tolerance, and reduced wavelength sensitivity---all favorable for a broadband passive reflector.
\begin{figure}[!t]
\centering
\includegraphics[width=3in]{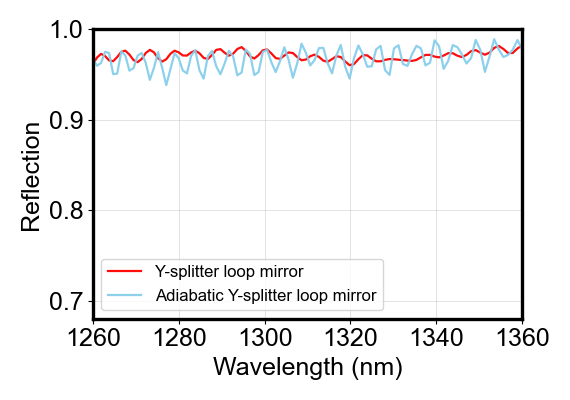}
\caption{FDTD simulated Reflection spectrum of Y-splitter loop mirror and adiabatic Y-splitter mirror.}
\label{fig5}
\end{figure}
The corresponding FDTD-simulated reflection spectra are plotted in Fig.\ref{fig5} over the $1260$--$1360~\mathrm{nm}$ wavelength range. Both loop-mirror designs maintain a broadband reflectivity of approximately $97\%$ across this O-band window, with bandwidths exceeding $100~\mathrm{nm}$, confirming their suitability as passive external reflectors for unidirectional MRL operation. Each spectrum shows a residual ripple associated with residual transmission and interference within the loop, and in these nominal simulations the two geometries reach comparable peak reflectivity, so the reflection magnitude alone does not strongly distinguish them. As discussed in the experimental section, the more relevant difference between the two splitters lies in their sensitivity to fabrication-induced deviations from an ideal $50{:}50$ split rather than in their nominal reflectivity. The subsequent experimental results examine how these reflector characteristics influence the threshold, slope efficiency, and spectral behavior of the complete laser cavity.

\subsection{DBR design}
To provide wavelength-selective feedback, a distributed Bragg reflector (DBR) was integrated at the end of the bus waveguide, as shown in Fig.\ref{fig4}(c). The grating section is formed in a $1~\mu\mathrm{m}$-wide silicon waveguide using a periodic corrugation with a fill factor of $50\%$, a grating period of $\Lambda = 626.353~\mathrm{nm}$, and a total of $160$ periods. The corrugation is defined over a lateral width of $4~\mu\mathrm{m}$ in the patterned region, while the etched portion of the DBR introduces an additional shallow etch depth of $40~\mathrm{nm}$ relative to the surrounding silicon rib waveguide. The corresponding FDTD-simulated field distribution is shown in Fig.\ref{fig4}(c), where the periodic longitudinal modulation of the optical field confirms the Bragg reflection established by the distributed grating. In contrast to the broadband loop-mirror reflectors discussed previously, the DBR provides spectrally selective feedback determined by its grating period, effective index contrast, and total number of periods.

\begin{figure}[!t]
\centering
\includegraphics[width=3in]{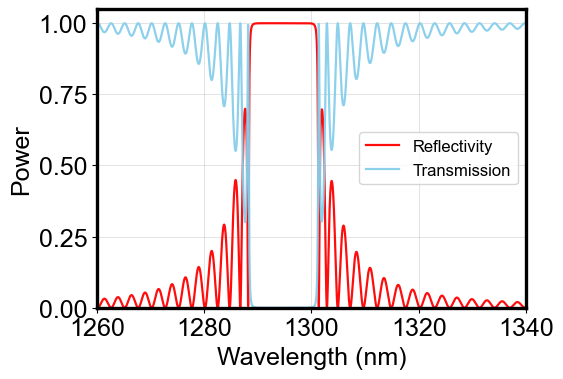}
\caption{Simulated reflection spectrum of the DBR, showing the high-reflectivity stopband near the Bragg wavelength.}
\label{fig7}
\end{figure}

The reflection spectrum of the DBR was calculated using the transfer matrix method (TMM), as shown in Fig.\ref{fig7}. The grating is formed by periodically alternating an unetched silicon waveguide section with effective index $n_{\mathrm{eff,high}} = 3.169$ and a shallow-etched section with effective index $n_{\mathrm{eff,low}} = 3.032$. The DBR operates as a third-order Bragg reflector, with a Bragg wavelength given approximately by $\lambda_B = 2 \bar{n}_{\mathrm{eff}} \Lambda / 3$, where $\bar{n}_{\mathrm{eff}} = (n_{\mathrm{eff,high}} + n_{\mathrm{eff,low}})/2$. This yields a center reflection wavelength near $1295~\mathrm{nm}$, in agreement with the high-reflectivity stopband shown in Fig.\ref{fig7}. The calculated spectrum exhibits a strong reflection plateau with a complementary transmission dip around the Bragg wavelength, confirming that the DBR provides wavelength-selective feedback near the target O-band lasing region. 

\section{Experimental characterization}

\subsection{Experimental characterization and results}

The light--current--voltage (LIV) characteristics were measured by current injection using a Keithley 2400 source measurement unit. Optical power from the left and right grating couplers was collected using cleaved SMF-28 fibers positioned at an angle of $7^\circ$. The collected optical power was measured with a Newport 2936-C power meter, while the optical spectra were simultaneously acquired through a $99/1$ directional coupler using a Yokogawa AQ6370E optical spectrum analyzer. The measured TE grating-coupler loss was on average $\sim -8.7~\mathrm{dB}$ per coupler at $1310~\mathrm{nm}$. During measurement, the $100~\mathrm{mm}$ wafer was vacuum-mounted onto a copper chuck and held at a controlled stage temperature of $25^\circ\mathrm{C}$. All reported LI responses were normalized to the measured grating-coupler losses. In addition, pre-bonded silicon waveguides with a width of $0.5~\mu\mathrm{m}$ were measured to have TE propagation losses of $\sim -22.1~\mathrm{dB/cm}$ at $1310~\mathrm{nm}$, as determined from cut-back test structures.

\begin{figure}[!t]
\centering
\includegraphics[width=3in]{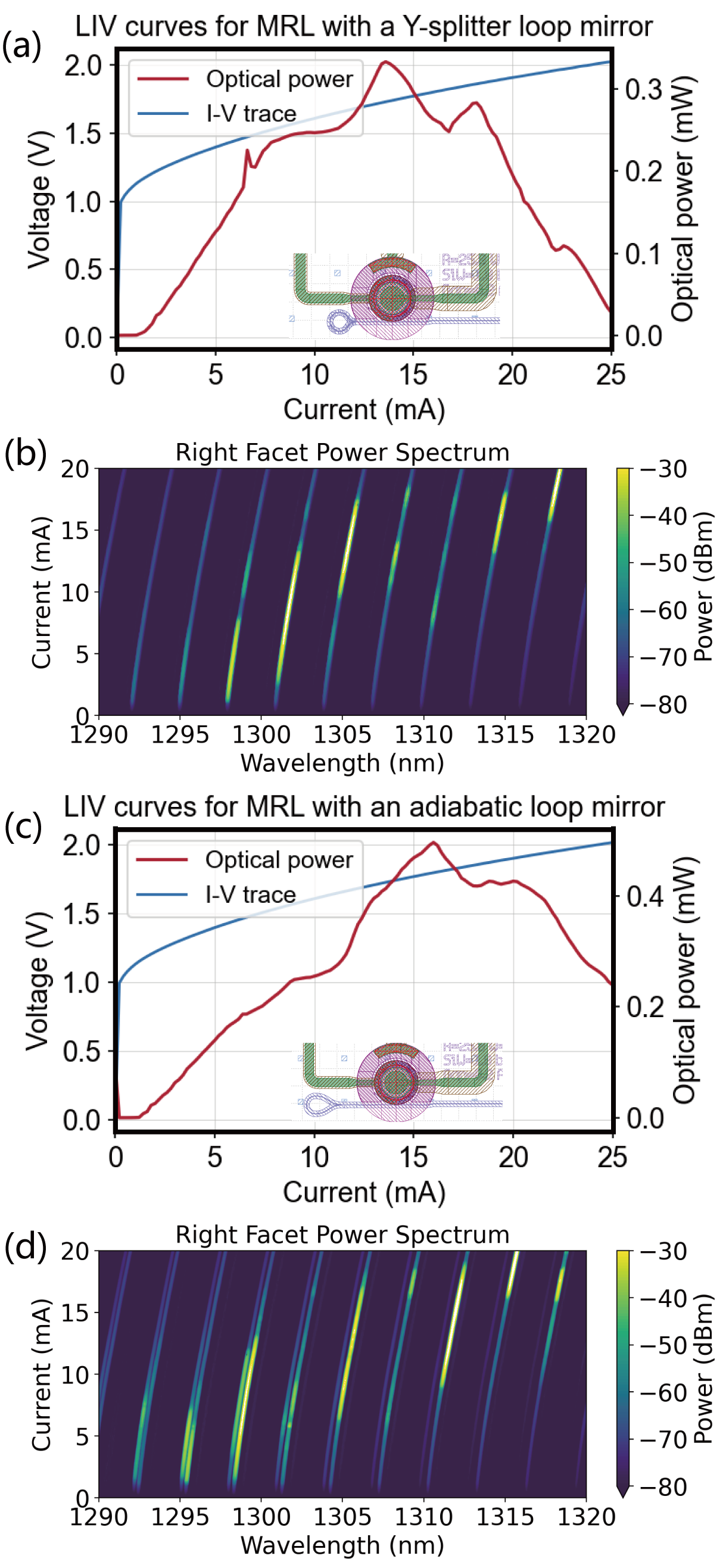}
\caption{Measured light–current–voltage (LIV) curves and optical spectra for (a), (b) the MRL with a Y-splitter loop mirror and (c), (d) the MRL with an adiabatic Y-splitter loop mirror. Insets in (a) and (c) show the layouts of the corresponding devices.}
\label{fig8}
\end{figure}

Figure~\ref{fig8}(a)--(d) shows the measured LIV characteristics and current-dependent optical spectra of the two loop-mirror-assisted devices. For the conventional Y-splitter loop mirror [Fig.~\ref{fig8}(a),(b)], lasing turns on at a low threshold, and because the loop mirror leaves only a single output port, the entire optical output is collected from the right facet. This confirms that the loop-mirror feedback effectively converts the intrinsically bidirectional microring into a single-ended emitter. Since the loop mirror is located on the bus waveguide outside the lasing ring, it leaves the ring cavity essentially unperturbed, so the threshold remains low and comparable to the baseline, while the single-facet output power is significantly higher than that of the bidirectional baseline. A similar directional behavior is observed for the adiabatic loop-mirror device [Fig.~\ref{fig8}(c),(d)], where the output is again collected from the right facet. Compared with the conventional Y-splitter loop mirror, the adiabatic reflector delivers a higher peak power and WPE. Because the two reflectors provide comparable nominal reflectivity ($\sim$97\%, Fig.~\ref{fig5}), we attribute this improvement to the superior fabrication tolerance of the adiabatic junction: its gradual mode evolution preserves a near-ideal $50{:}50$ power split even under lithographic and etch variations, whereas the abrupt conventional Y-junction is far more sensitive to such variations. When the conventional splitter deviates from a $50{:}50$ split, the two arm fields no longer recombine fully at the junction, lowering the effective loop reflectivity and the recovered single-facet power. This robustness comes at the expense of a larger device footprint owing to the long adiabatic transition. Taken together, these two devices show that loop-mirror feedback can enforce single-ended emission at a low threshold, while the splitter geometry sets a trade-off between fabrication tolerance and device footprint.

\begin{table*}[t]
\caption{Experimental summary of uni-directional MRL designs.}
\label{tab:summary_uni_mrl}
\centering
\renewcommand{\arraystretch}{1.08}
\begin{tabular*}{\textwidth}{@{\extracolsep{\fill}}lcccc@{}}
\toprule
Uni-directional Design & Threshold current & Peak power & WPE (peak value) & $P_{\mathrm{CW}}/P_{\mathrm{CCW}}$ (peak WPE) \\
\midrule
None (Baseline MRL) & 0.80 mA & 0.159 mW & 0.612\% & $-1.92$ dB \\
Y-splitter loop mirror & 0.99 mA & 0.237 mW & 1.047\% & $\infty^{*}$ \\
Adiabatic Y-splitter loop mirror & 1.12 mA & 0.258 mW  & 1.119\% & $\infty^{*}$ \\
DBR reflector & 0.79 mA & 0.200 mW & 0.805\% & 27.65 dB \\
\bottomrule
\end{tabular*}

\vspace{3pt}
\parbox{\textwidth}{For each of the three external-reflector designs (Y-splitter, adiabatic Y-splitter, and DBR), the reported threshold current, peak power, and peak WPE are averaged over four measured devices. Peak wall-plug efficiency (WPE) is taken at the peak-power operating point using grating-coupler-normalized output. The directionality metric $P_{\mathrm{CW}}/P_{\mathrm{CCW}}$ is evaluated at the current of maximum WPE. The baseline MRL without uni-directional feedback structures is listed as a single representative device. $^{*}$Lasers have only one output facet.}
\end{table*}

\begin{figure}[!t]
\centering
\includegraphics[width=3in]{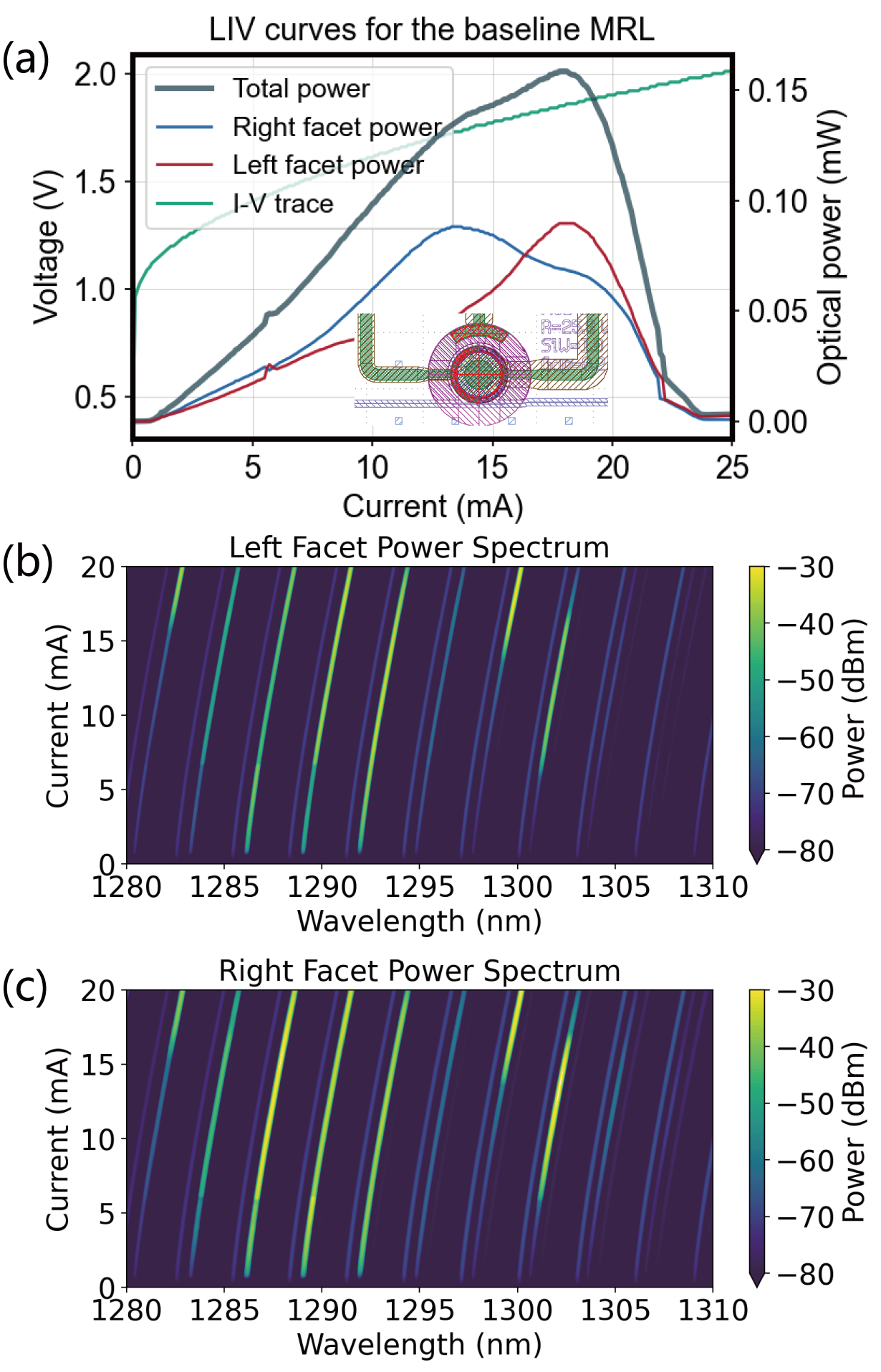}
\caption{Measured light–current–voltage (LIV) characteristics and current-dependent optical spectra of the baseline MRL without reflective feedback. (a) LIV curves; (b), (c) optical power spectra versus injection current collected from the left and right facets, respectively. Inset in (a): baseline device layout.}
\label{fig9}
\end{figure}

\begin{figure}[!t]
\centering
\includegraphics[width=3in]{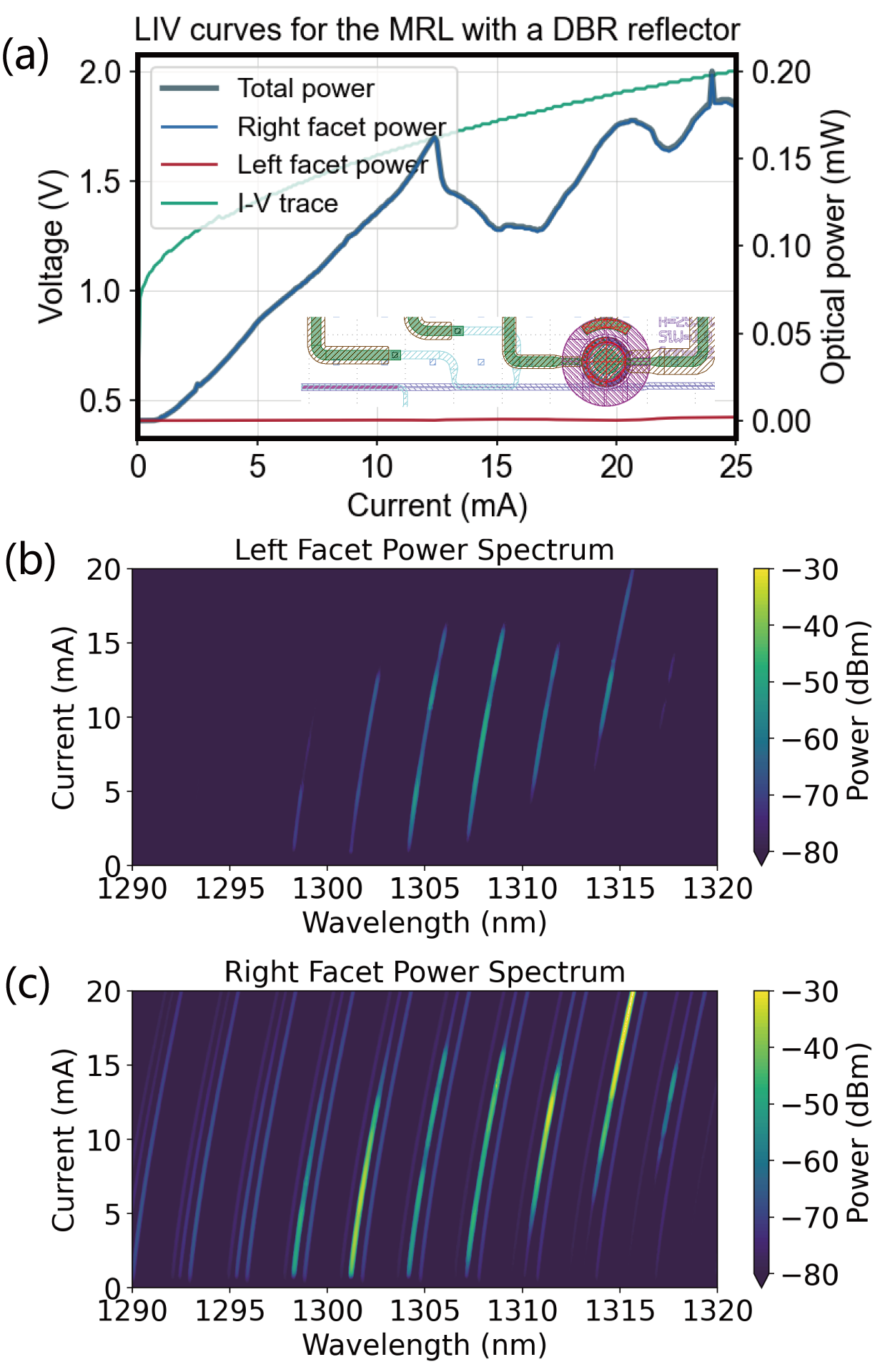}
\caption{Measured light–current–voltage (LIV) characteristics and current-dependent optical spectra of the MRL with an integrated DBR reflector. (a) LIV curves; (b), (c) optical power spectra versus injection current collected from the left and right facets, respectively. Inset in (a): device layout with the DBR integrated at one end of the bus waveguide.}
\label{fig10}
\end{figure}

The baseline and DBR-assisted devices are compared in Fig.~\ref{fig9} and Fig.~\ref{fig10}. For the baseline reflector-free microring [Fig.~\ref{fig9}(a)--(c)], the left- and right-facet powers remain comparable across the measured current range, and the optical spectra recorded from the two facets exhibit similar mode branches, consistent with the expected bidirectional lasing of the unperturbed microring cavity, for which no directional feedback is present. The spectra consist of a comb of longitudinal modes spaced by approximately the ring free-spectral range ($\sim 3~\mathrm{nm}$), each of which red-shifts continuously with increasing current as a result of self-heating. The weaker secondary peaks that appear within each free-spectral range form a second comb that we attribute to a higher-order transverse mode of the ring. Finite-difference eigenmode (FDE) analysis of the bent cavity cross-section supports this assignment: the cavity supports two low-loss TE modes, the fundamental and the first higher-order radial mode, with group indices $n_g \approx 3.67$ and $3.74$ and corresponding free-spectral ranges of $2.99$ and $2.94~\mathrm{nm}$, respectively, in good agreement with the two measured free-spectral ranges ($\sim 3.00$ and $\sim 2.97~\mathrm{nm}$). The higher-order mode further exhibits an approximately fivefold larger simulated radiation loss ($0.99$ versus $0.21~\mathrm{dB/cm}$) and a greater spatial overlap with the cavity sidewalls, which raises its surface-recombination and optical scattering losses and hence its threshold; the fundamental mode consequently reaches lasing first and, through gain clamping, holds the higher-order mode below its own threshold, so that the latter appears only as the weak secondary branches observed here. In contrast, once the DBR reflector is introduced [Fig.~\ref{fig10}(a)--(c)], the output becomes strongly asymmetric: one facet carries the dominant optical power, whereas the opposite facet is substantially suppressed over most of the operating range. At the same time, the threshold remains low and comparable to the baseline, as expected since the DBR is integrated on the bus waveguide outside the lasing ring and does not perturb the ring cavity. For the baseline cavity the two facets carry comparable mode combs, whereas for the DBR device the suppressed facet retains only weak residual branches while the dominant facet preserves the full comb, confirming that the DBR biases the microring toward directional emission. The principal difference between the two devices is therefore the balance of power between the facets. Overall, both loop mirrors and DBRs break the bidirectional symmetry of the baseline MRL while preserving its low threshold; the choice of reflector then trades off the completeness of the directional suppression, the output power and efficiency, and the device footprint.

Table~\ref{tab:summary_uni_mrl} summarizes the measured performance of the baseline and reflector-assisted uni-directional MRLs. For each of the three external-reflector designs, the threshold current, peak power, and peak WPE are averaged over four measured devices. All four designs maintain a comparable low threshold current of around 1 mA; since the reflectors are integrated on the bus waveguide outside the lasing ring, they do not modify the ring cavity, and the small device-to-device spread in threshold reflects fabrication variation rather than a systematic effect of the reflector. The baseline device exhibits nearly balanced bidirectional emission, as indicated by the low directionality ratio of $P_{\mathrm{CW}}/P_{\mathrm{CCW}} = -1.92~\mathrm{dB}$, with a peak power of $0.159~\mathrm{mW}$ and a peak WPE of $0.612\%$. All three reflector-assisted designs break this bidirectional symmetry and, by concentrating the emission into a single facet, reach a higher peak power and WPE than the baseline cavity. Both Y-splitter loop mirrors achieve effectively single-ended emission; between them, the adiabatic loop mirror delivers a higher peak power ($0.258~\mathrm{mW}$) and WPE ($1.119\%$) than the conventional Y-splitter ($0.237~\mathrm{mW}$ and $1.047\%$), consistent with its greater tolerance to fabrication-induced deviations from an ideal $50{:}50$ split, at the expense of a larger device footprint due to its long adiabatic transition. The DBR-assisted device provides a strong directionality ratio of $27.65~\mathrm{dB}$ with a peak power of $0.200~\mathrm{mW}$ and a WPE of $0.805\%$. These results show that all of the external reflectors preserve the low threshold of the baseline MRL and enhance its single-facet output, while the specific reflector geometry sets the trade-off among output power, efficiency, directionality, and device footprint.

\subsection{High-Speed Measurements}

\begin{figure}[!t]
    \centering
    \includegraphics[width=\columnwidth]{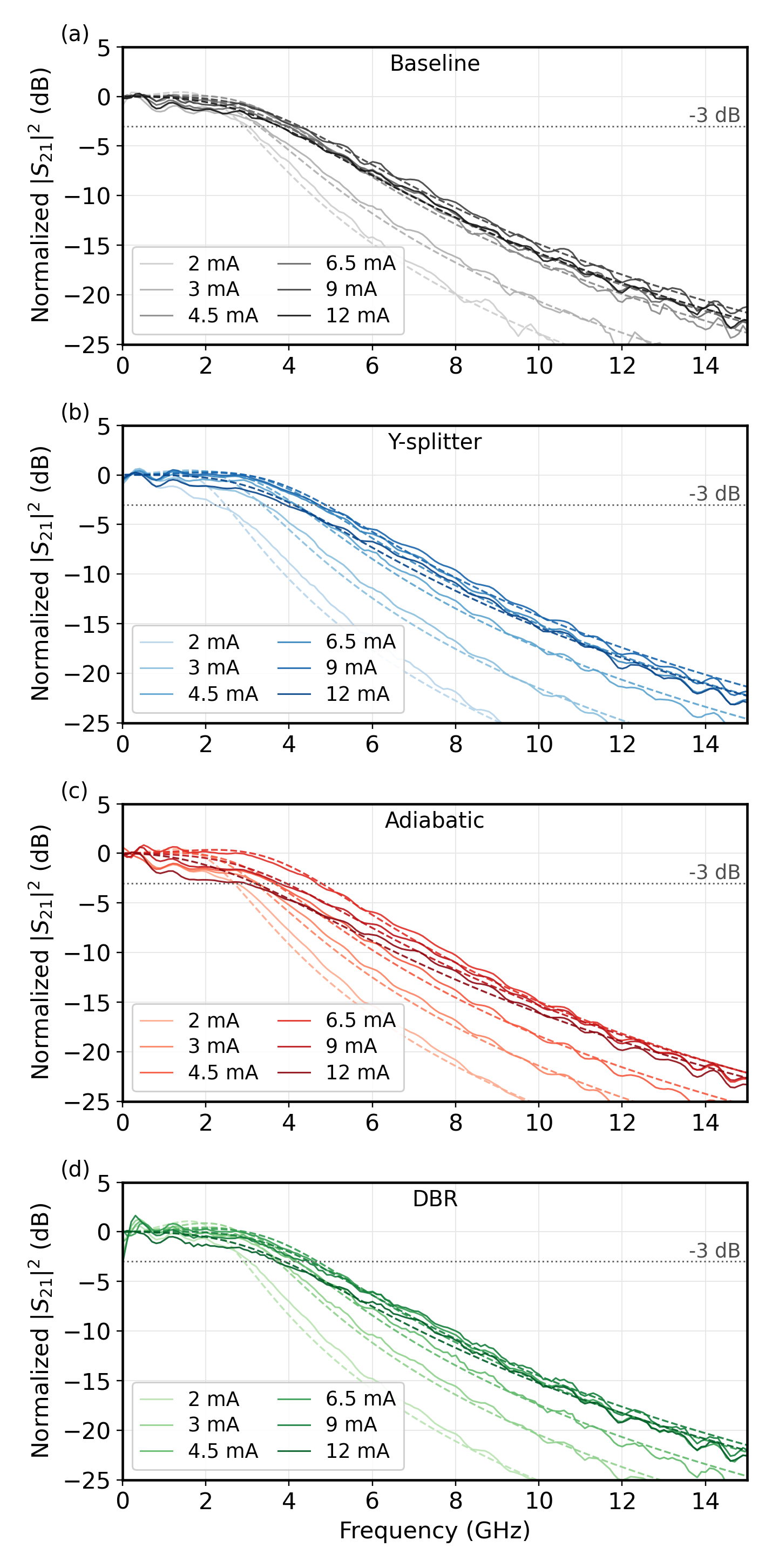}
    \caption{Measured normalized $|S_{21}|^2$ modulation responses (solid lines,
    0.04--15~GHz) of the four reflector configurations biased from 2.0 to
    12.0~mA: (a)~bidirectional baseline, (b)~Y-splitter loop mirror,
    (c)~adiabatic Y-splitter loop mirror, and (d)~DBR. Dashed lines are the
    corresponding three-pole transfer-function fits, and the dotted horizontal
    line marks the $-3$~dB level.}
    \label{fig11}
\end{figure}

In addition to the directionality and ultra-low-threshold operation established above, we investigated whether the choice of unidirectional reflector affects the modulation bandwidth of the MRL. Small-signal $S$-parameter characterization was performed using an Agilent E8363B vector network analyzer (VNA) with a 50~GHz bandwidth and a signal--ground (SG) RF probe, with the sample mounted on a temperature-controlled stage held at $T = 25\,^\circ$C. The measurements were carried out on the four reflector configurations introduced above---the
bidirectional baseline (no reflector), the Y-splitter loop mirror, the adiabatic Y-splitter loop mirror, and the DBR---each characterized over a bias range of 2.0 to 12.0~mA. Figs.~\ref{fig11}(a)--(d) show the resulting normalized $|S_{21}|^2$ responses, measured from 0.04 to 15~GHz, for the baseline, Y-splitter, adiabatic, and DBR devices, respectively; the experimental responses (solid lines) are well fitted by the standard three-pole modulation transfer function (dashed lines)~\cite{ref32}, from which the relaxation oscillation frequency $f_r$ and damping rate $\gamma$ are extracted at each bias and the 3-dB bandwidth
$f_\mathrm{3dB}$ is read directly from the measured data.

\begin{figure}[!t]
    \centering
    \includegraphics[width=\columnwidth]{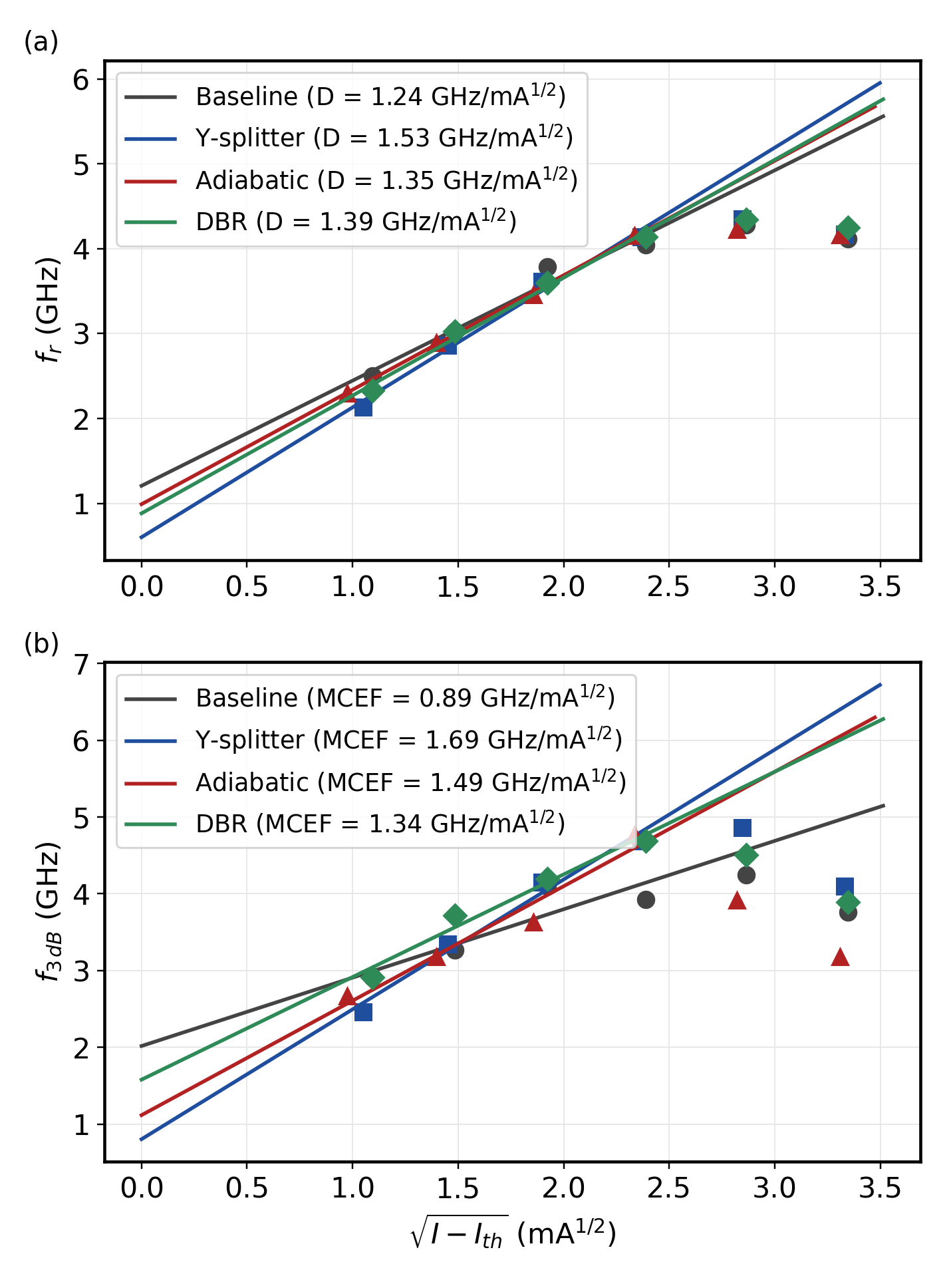}
    \caption{Extracted (a) relaxation-oscillation frequency $f_r$ and (b) 3-dB
    bandwidth $f_\mathrm{3dB}$ as a function of $\sqrt{I-I_\mathrm{th}}$ for the
    four reflector configurations (baseline, Y-splitter, adiabatic, and DBR).
    Solid lines are linear fits to the low-bias data, yielding the $D$-factor in
    (a) and the modulation current efficiency factor (MCEF) in (b).}
    \label{fig12}
\end{figure}

Figs.~\ref{fig12}(a) and (b) plot the extracted $f_r$ and $f_\mathrm{3dB}$versus $\sqrt{I-I_\mathrm{th}}$ for the four configurations, with linear fits to the first four bias points yielding the $D$-factor and the modulation current efficiency factor (MCEF), respectively. The fitted $D$-factors are 1.24, 1.53, 1.35, and 1.39~GHz/mA$^{1/2}$ for the baseline, Y-splitter, adiabatic, and DBR devices, with corresponding MCEF values of 0.89, 1.69, 1.49, and 1.34~GHz/mA$^{1/2}$. The relaxation-oscillation data nearly overlap across all four devices and the $D$-factors fall within a narrow range with no systematic
dependence on reflector type, indicating that the intrinsic modulation efficiency is essentially unaffected by the addition of an external reflector. Although the MCEF values exhibit somewhat larger scatter, the resulting 3-dB bandwidths remain comparable across all four devices, reaching maxima of approximately 4--5~GHz with
no degradation attributable to any reflector design. Taken together, these results indicate that the unidirectional reflectors impose no measurable penalty on the modulation bandwidth, demonstrating that single-direction operation can be achieved without compromising the high-speed performance of the MRL.

\begin{figure}[!t]
    \centering
    \includegraphics[width=\columnwidth]{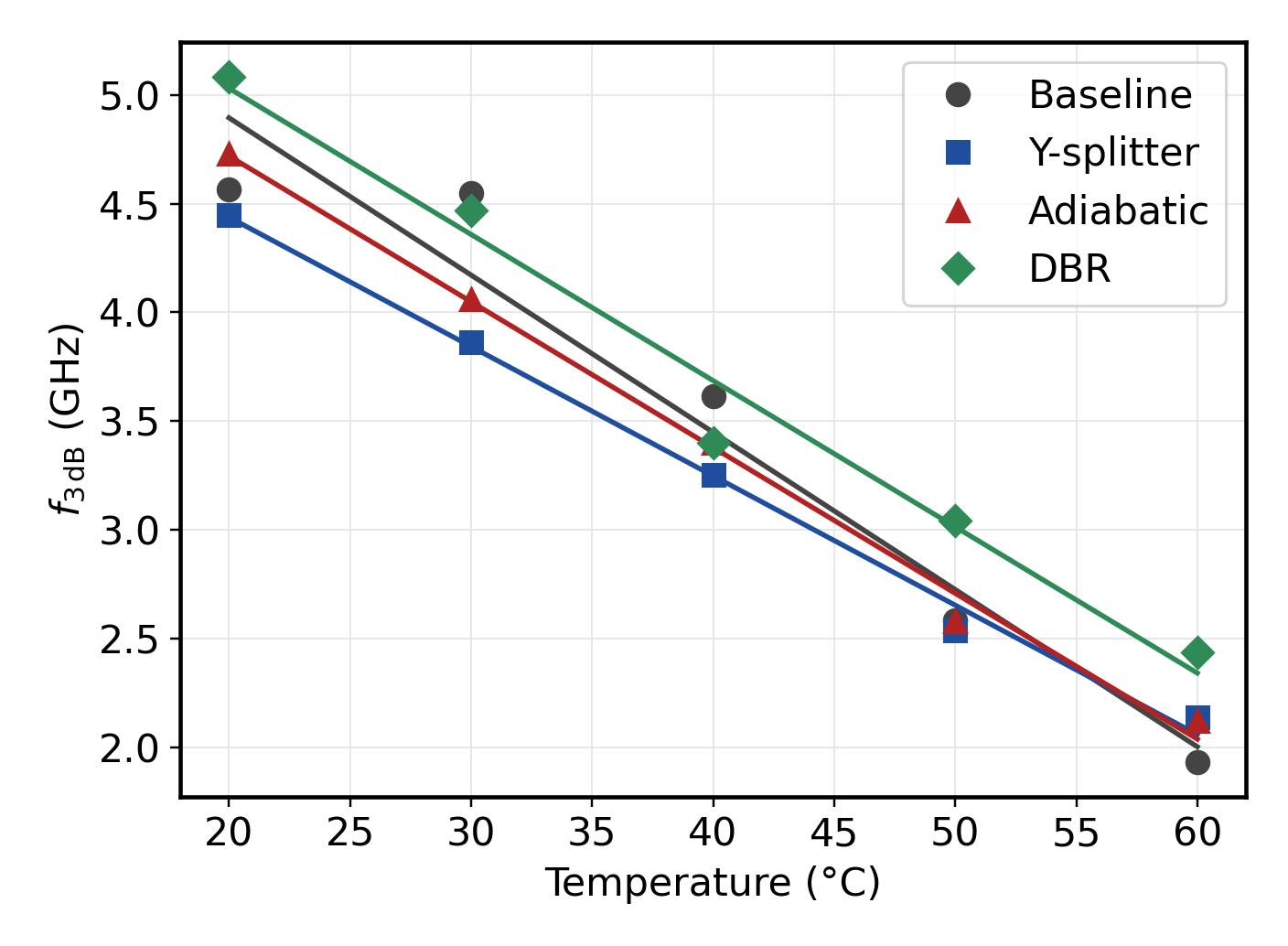}
    \caption{Extracted 3-dB bandwidth $f_\mathrm{3dB}$ versus stage temperature
    for the four reflector configurations (baseline, Y-splitter, adiabatic, and
    DBR), measured at a fixed bias current of 6.5~mA. Solid lines are linear fits
    to the data.}
    \label{fig13}
\end{figure}

To assess the thermal robustness of the modulation response, the small-signal $S$-parameter measurements were repeated at stage temperatures from 20 to 60~$^\circ$C in 10~$^\circ$C increments, at a fixed bias current of 6.5~mA for all four devices. As shown in Fig.~\ref{fig13}, the extracted $f_\mathrm{3dB}$ decreases approximately linearly with temperature for every configuration, from 4.5--5.1~GHz at 20~$^\circ$C to 1.9--2.4~GHz at 60~$^\circ$C. Linear fits yield thermal degradation rates of $df_\mathrm{3dB}/dT = -0.072$, $-0.059$, $-0.067$, and $-0.067$~GHz/$^\circ$C for the baseline, Y-splitter, adiabatic, and DBR devices, corresponding to fractional rates of $-1.48$, $-1.34$, $-1.42$, and $-1.34$~\%/$^\circ$C, respectively. These rates are comparable across all four configurations and show no systematic dependence on the reflector type, indicating that the thermal degradation of the modulation bandwidth is governed by the temperature dependence of the common QD gain medium rather than by the output reflector. Consistent with the room-temperature results, the unidirectional reflectors therefore have negligible impact on both the magnitude and the thermal
robustness of the modulation bandwidth.

\section{Conclusion}
We have demonstrated and systematically compared external-reflector-assisted hybrid III--V/Si quantum-dot microring lasers that achieve stable unidirectional emission while preserving ultra-low threshold operation. Using an extended coupled-mode-theory rate-equation model, we showed that integrating a passive reflector at the end of the bus waveguide breaks the CW/CCW symmetry through an asymmetric inter-modal coupling term, biasing the cavity toward a single circulation direction without perturbing the ring itself. Three passive feedback architectures---a conventional Y-splitter loop mirror, an adiabatic Y-splitter loop mirror, and a distributed Bragg reflector (DBR)---were fabricated and benchmarked against a reflector-free bidirectional baseline. All devices retained ultra-low thresholds of 0.79--1.12~mA, and each reflector enforced effectively single-ended emission, confirming that passive external feedback can convert the intrinsically bidirectional MRL into a directional source. The reflector geometry, however, governs the trade-off between directionality and laser performance: the loop mirrors provide clean single-facet routing, and the two splitter geometries trade fabrication tolerance against footprint. Although the two loop mirrors provide comparable nominal reflectivity, the abrupt conventional Y-splitter is more sensitive to fabrication-induced deviations from a $50{:}50$ split and therefore exhibits a higher effective insertion loss and lower output power, whereas the adiabatic Y-splitter loop mirror preserves a near-ideal $50{:}50$ split---and correspondingly higher output power---at the expense of a larger footprint owing to its 20~$\mu$m adiabatic transition; among all configurations, the adiabatic loop mirror yields the highest single-facet peak power and wall-plug efficiency. By contrast, the DBR preserves the lowest threshold of the cavity while delivering the highest measured directional isolation, 27.65~dB.

Small-signal modulation measurements further showed that the unidirectional reflectors impose no measurable penalty on the high-speed performance of the laser: the extracted $D$-factors and 3-dB bandwidths were comparable across all four configurations, reaching maxima of approximately 4--5~GHz, and the thermal degradation of the bandwidth was governed by the shared QD gain medium rather than by the reflector. Taken together, these results establish passive external feedback as a practical, low-invasiveness route to controlled unidirectional emission that preserves the intrinsic low-threshold and high-speed advantages of hybrid III--V/Si QD microring lasers, offering useful design guidance for their integration into energy-efficient, DWDM-scale optical interconnects.

\vfill
\end{document}